\documentclass[twocolumn,superscriptaddress,prl]{revtex4}
\usepackage{color}
\usepackage{xspace}
\usepackage{amsmath}
\usepackage{amssymb}
\usepackage{amsbsy}
\usepackage{graphicx}
 \usepackage{bm}
 \usepackage{float}

\usepackage{relsize}

\newcommand{\br}{\boldsymbol{\rho}}

\newcommand{\x}{\mathbf{r}}

\begin{document}

\title{Scaling of Fluctuations in a Trapped Binary Condensate}

\author{R.~N.~Bisset}
\affiliation{Center for Nonlinear Studies and Theoretical Division, Los Alamos National Laboratory, Los Alamos, New Mexico 87545, USA}
\author{R.~M.~Wilson}
\affiliation{Department of Physics, The United States Naval Academy, Annapolis, MD 21402, USA}
\affiliation{Joint Quantum Institute, National Institute of Standards and Technology and University of Maryland, Gaithersburg, MD 20899, USA}
\author{C.~Ticknor}
\affiliation{Center for Nonlinear Studies and Theoretical Division, Los Alamos National Laboratory, Los Alamos, New Mexico 87545, USA}

\begin{abstract}

We demonstrate that measurements of number fluctuations within finite cells provide a direct means to study susceptibility scaling in a trapped two-component Bose-Einstein condensate.
This system supports a second-order phase transition between miscible (co-spatial) and immiscible (symmetry-broken) states that is driven by a diverging susceptibility to magnetic fluctuations.
As the transition is approached from the miscible side the magnetic susceptibility is found to depend strongly on the geometry and orientation of the observation cell.
However, a scaling exponent consistent with that for the homogenous gas ($\gamma = 1$) can be recovered, for all cells considered, as long as the fit excludes the region in the immediate vicinity of the critical point.
As the transition is approached from the immiscible side, the magnetic fluctuations exhibit a non-trivial scaling exponent $\gamma \simeq 1.30$.
Interestingly, on both sides of the transition, we find it best to extract the exponents using an observation cell that encompasses half of the trapped system. This implies that relatively low-resolution \emph{in situ} imaging will be sufficient for the investigation of these exponents.
We also investigate the gap energy and find exponents $\nu z$ = 0.505 on the miscible side and, unexpectedly, $\nu z$ = 0.60(3) for the immiscible phase.

\end{abstract}

\maketitle

\section{Introduction}

Ultracold quantum gases play an important role in our understanding of phase transitions; they provide us with a number of experimental controls and observational tools that are not available in conventional condensed matter systems.  For example, modern experiments with laser-cooled atoms have demonstrated thermal phase transitions such as Bose-Einstein condensation (BEC) \cite{Anderson1995,Bradley1995,Davis1995,Ensher1996,Gerbier2004,Tammuz2011} and the Berezinkskii-Kosterlitz-Thouless transition in two-dimensional (2D) superfluids~\cite{Hadzibabic2006,Clade2009,Tung2010,Hung2011}, and quantum phase transitions such as the Mott insulator/superfluid transition \cite{Greiner2002,Bakr2009,Sherson2010,Bakr2010} and Dicke super-radiant self-organization~\cite{Baumann2010,Brennecke2013,Mottl2012}.  Recently, a great deal of interest has been generated by quantum gases of bosons with spin or pseudo-spin degrees of freedom, including spinor~\cite{Kawaguchi2012,StamperKurn2013} and spin-orbit coupled BECs~\cite{Lin2011,Zhang2012,Parker2013}, which are ideal for studying a variety of magnetic and topological phase transitions in quantum-degenerate matter.

Perhaps the most simple, yet robust quantum Bose gas with a pseudo-spin (1/2) degree of freedom is the two-component, or binary BEC.  At ultracold temperatures, this system supports a second-order phase transition between miscible and immiscible states as the interaction strength between the components ($g_{12}$) is tuned across a critical threshold $g_c$, reminiscent of a para-to-ferromagnetic Ising transition.  
To date, a great deal of experimental~\cite{Papp2008,Hall1998B} and theoretical~\cite{Esry1997,Riboli2002,Svidzinsky2003,Trippenbach2000,Timmermans1998,Ao1998,Ao2000,Alexandrov2002,Santamore2012,Oles2008,Zhan2014,Pattinson2013,Roy2014} work has been dedicated to understanding these states and the dynamics of the transition between them.

Binary condensates exhibit qualitatively different behavior, if the population of each component is required to be individually conserved, depending whether the system is trapped or homogenous (infinite).
For the trapped system, in the immiscible phase, the separated components meet at a well-defined interface where the components overlap (see Fig.~\ref{Fig:Fig1} (b) and Fig.~\ref{Fig:FlucScaling} insets). In contrast, the infinite homogenous system is somewhat pathological with the interface boundary being difficult to define.
We consider the case where the population of each component is individually conserved, and focus on the experimentally realistic situation of an oblate trapping potential, much like a pancake.
As we shall see, a further consideration for the trapped system is that large length-scale fluctuations are dominated by a finite number of excitations, in contrast to the continuum exhibited by the infinite system.


A number of groups have performed theoretical studies of quenches across the miscible-immiscible transition for both simple \cite{Mertes2007,Anderson2009,Kasamatsu2004,Ronen2008,Hofmann2014} and coherently coupled \cite{Nicklas2014,Sabbatini2011} binary condensates, and have demonstrated, for example, a Kibble-Zurek scaling of domain formation~\cite{Sabbatini2011} and long-time coarsening behavior of the ``ferromagnetic'' domains~\cite{Hofmann2014}.  These studies were motived in part by the fact that ultracold atomic systems equilibrate on relatively long time scales, allowing for resolved measurements of their dynamics.
Modern methods for \textit{in situ} imaging, however, allow for the direct measurement of number fluctuations, and thus for statistical studies of equilibrium states that were previously unobtainable \cite{Bakr2009}.
Experimental \textit{in situ} investigations of atom-number fluctuations have already probed quantum fluctuations in one dimension \cite{Jacqmin2011}, the universality of the Berezinkskii-Kosterlitz-Thouless transition \cite{Hung2011} and the Hanbury Brown-Twiss effect \cite{Blumkin2013} in Bose gases.
Furthermore, recent Theoretical studies pave the way for experiments to utilize number fluctuations to investigate spinor condensates \cite{Symes2014} and roton excitations in dipolar condensates \cite{Bisset2013B,Bisset2013,Klawunn2011}.

In this paper, we study the equilibrium properties of a binary condensate near the miscible-immiscible transition threshold at very low but finite temperature, and
show how measurements of number fluctuations can reveal the scaling exponent associated with the diverging susceptibility to pseudo-spin, or magnetic fluctuations.
We numerically calculate the magnetic fluctuations within finite observation cells.
Experimentally, the observation cells may be formed either by considering individual imaging pixels or by combining pixels to form larger cells, and fluctuation statistics can be obtained by repeated \emph{in situ} imaging.
We extract scaling exponents for the the gap energy $\Delta$ and the magnetic susceptibility $\chi$ as~\cite{SachdevBook}
\begin{align}
\Delta \sim |\delta|^{\nu z},~\chi \sim |\delta|^{-\gamma} \label{Eq:CritScale}
 \end{align}
where $\delta \equiv 1 - g_{12} / g_c$, and show how the latter exhibits a strong dependence on the cell orientation and geometry.
The scaling of the homogeneous binary BEC ($\gamma = 1$) can be recovered in the harmonically trapped system as long as the fitting region excludes the immediate vicinity of the critical point or, perhaps counterintuitively, by choosing a large cell that encapsulates half the system.
We also study the miscible-immiscible transition as it is approached from the immiscible side, and find a non-trivial scaling exponent $\gamma \simeq 1.30$.  Finally, we compare our results with a local density approximation (LDA), and find qualitative agreement only for very specific orientations of the observation cell.

\section{Formalism}

We consider an ultracold Bose gas with two distinguishable components that have equal mass $m$ and are trapped by the same harmonic potential $V(\x) = m(\omega_x^2 x^2 + \omega_y^2 y^2 + \omega_z^2 z^2)/2$, where $\omega_i$ are the trapping frequencies along each direction.  Additionally, we consider the case where the confinement along the $z$-direction is tight ($\omega_z \gg \omega_x,\omega_y$) so the axial degree of freedom is effectively frozen out. This assumption is accurate if the temperature $k_BT/\hbar\omega_z\ll 1$ and chemical potential $\mu/\hbar\omega_z\ll 1$ are relatively small.
The oblate geometry is ideal for experimental measurements of number fluctuations, which can be made via \emph{in situ} column density imaging~\cite{Hung2011}.
  Analytically integrating over the $z$-coordinate then allows us to work in the quasi-two dimensional (quasi-2D) regime with the spatial coordinates $\br = \{x, y\}$.  The two-body scattering is still three dimensional (3D), however, provided that $a_{ij}\ll a_z$, where $a_{ij}$ is the 3D $s$-wave scattering length between components $i$ and $j$, and $a_z = \sqrt{\hbar/m\omega_z}$~\cite{Petrov00}.  In the weakly interacting limit, the condensate order parameters $\psi_k(\boldsymbol\rho) = \langle \hat{\Psi}_k(\br) \rangle$, where $k=1,2$, are solutions of the coupled Gross-Pitaevskii equations (GPEs)\cite{Esry1997,Pu1998,Ho1996},
\begin{equation}
\left[\frac{-\hbar^2{\nabla}^2}{2m} + V(\br) +\sum_i g_{ki} n_i^0(\br) \right]\psi_k(\br) = \mu_k\psi_k(\br), \label{Eq:GPE}
\end{equation}
with $n_i^0(\br) = |\psi_i(\br)|^2$ being the unit-normalized areal density of component $i$, $g_{ij} = N\sqrt{8\pi}\hbar^2 a_{ij}/ma_z$ and $N$ is the number of atoms per component.  We obtain the condensate excitations by linearizing the coupled Eqs.~(\ref{Eq:GPE}) about the stationary solutions $\psi_k(\br)$ and solving the resulting Bogoliubov de Gennes (BdG) equations~\cite{Ticknor2013,Pu1998,Pu1998A,Timmermans1998}.

In the thermodynamic limit, the fluctuation-dissipation theorem relates the magnetic susceptibility $\chi=2\partial(N_1-N_2)/\partial(\mu_1 - \mu_2)_T$ to the magnetic number fluctuations $\delta M^2$ as \cite{Recati2011,Seo2011}
\begin{equation}
\frac{ k_BT}{N_\sigma} \chi = \frac{\delta M^2}{N_\sigma} = \frac{\langle [\hat{M}_\sigma - \langle \hat{M}_\sigma\rangle ]^2 \rangle}{N_\sigma} , \label{Eq:MagFluc}
\end{equation}
where $\hat M_\sigma = \int_\sigma d \br [\hat{n}_1(\br )-\hat{n}_2(\br)]$ is the magnetization operator and $N_\sigma = \langle \hat{N}_\sigma \rangle = \int_\sigma d \br \langle \hat{n}_1(\br)+\hat{n}_2(\br) \rangle$ is the total particle number in the cell $\sigma$, $\mu_k$ is the chemical potential, and $\hat{n}_k(\br) = \hat\Psi_k^\dagger(\br) \hat\Psi_k(\br) $ is the density operator for component $k$.  At the level of Bogoliubov theory we evaluate $\delta M^2$ at quadratic order in the field fluctuations $\hat{\Psi}_k(\br) - \langle \hat{\Psi}_k(\br) \rangle$ and neglect higher order contributions which are vanishingly small away from the critical point. Bogoliubov theory is expected to give good quanitative results in the ultracold dilute regime that we consider here except in the vicinity of the critical point where interactions between excitations become important.
For detailed discussions of density fluctuations in condensed systems, see Refs.~\cite{Bisset2013B,Bisset2013,Klawunn2011}.

In practice, we solve the coupled GPEs (Eq.~(\ref{Eq:GPE})) and the BdG equations with a basis of ideal harmonic oscillator modes beneath the single-particle energy cutoff $E_{\mathrm{cut}}=100\hbar\omega_x$.  We use 8800 modes when evaluating Eq.~(\ref{Eq:MagFluc}) in the normally ordered form, which produces a converged result \footnote{We correct for the error introduced by such an approach, i.e.~that  \unexpanded{$\langle \psi_k^*(\x)\psi_k^*(\x)\psi_k(\x)\psi_k(\x)\rangle = (n_k^0)^2$} instead of $n_k^0(n_k^0-1)$. In practice this amounts to correcting the poissonian fluctuation contributions i.e.~$N_{\sigma k}^0 \to N_{\sigma k}^0 - (N_{\sigma k}^0)^2/N_k^0$ where $N_{\sigma k}^0=\int_\sigma n_k^0(\x)d^3\x$ and $N_k^0=\int n_k^0(\x)d^3\x$.}.

\section{Results}

We consider a system with interaction strengths $g_{11} = g_{22} \equiv g = 500\hbar^2/m$ in a slightly asymmetric trap with $\omega_y/\omega_x = 1.1$ to avoid the ambiguity of choosing a boundary axis. For this system, the critical interspecies interaction strength, which defines the threshold of the miscible-immiscible transition, is $g_c = 1.0072 (8) g $~\footnote{The uncertainty of the critical point occurs due to a small region of coexistence exhibiting both a stable miscible and an immiscible solution.}, where the deviation from unity is due to finite size effects.
\begin{figure} 
\begin{center}
\includegraphics[width=3.3in]{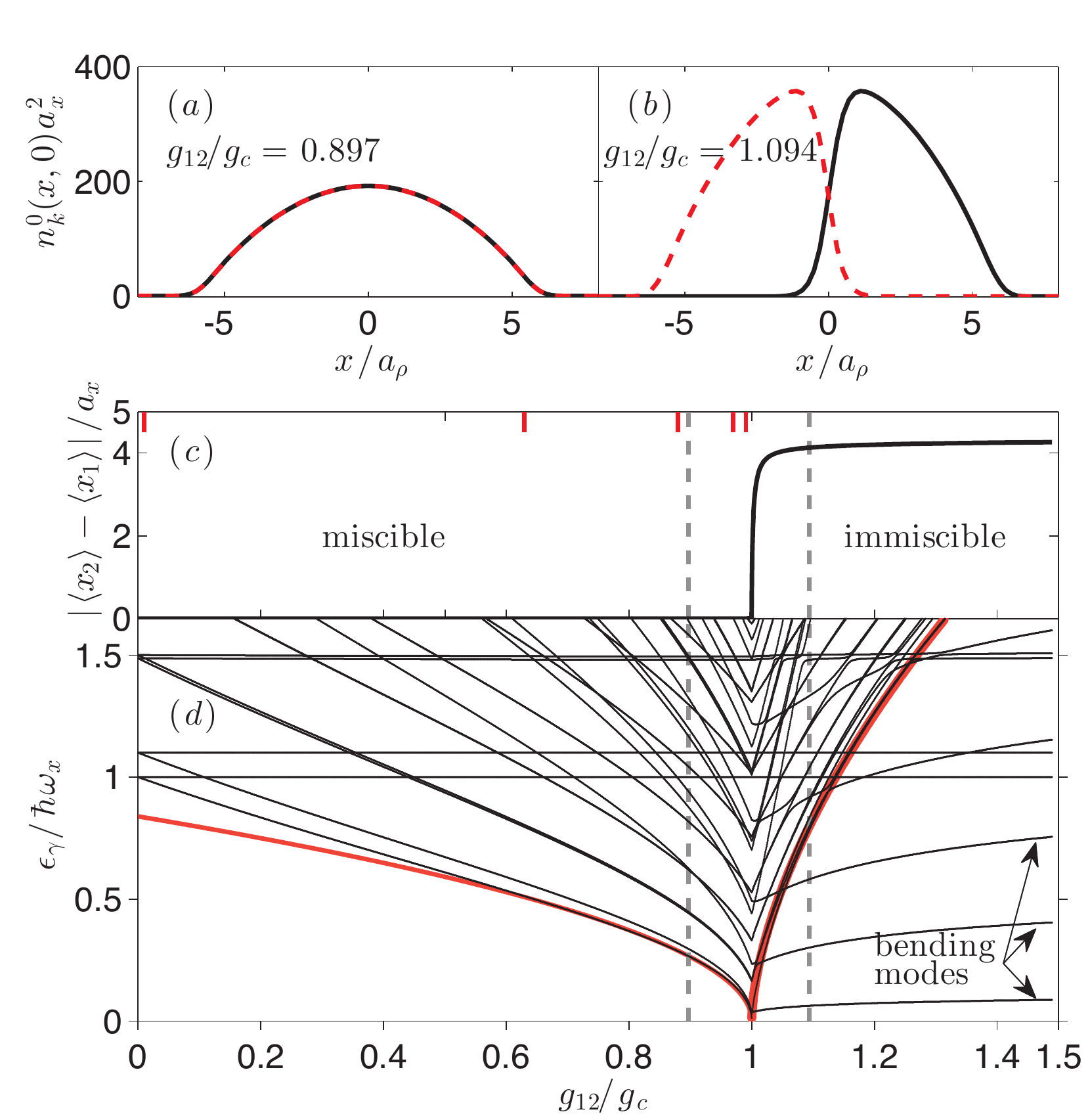} 
\caption{(color online) Transition properties. Condensate densities of component 1 (dashed) and 2 (solid) in miscible (a) and immiscible (b) phases. (c) Component separation in the $x$-direction, where $\langle x_k \rangle = \int d\br \, x \, n_k^0(\br)$. Short red (gray) vertical lines mark the five smallest $g_{12}/g_c$ plotted in Fig.~\ref{Fig:FlucVsPos}. (d) Bogoliubov energy spectrum.
The red (gray) solid curves are scaling fits with $\nu z=0.505$ for the miscible and $\nu z=0.60(3)$ for the immiscible region (see text). The vertical dashed lines show the interaction parameters for (a) and (b). $a_x = \sqrt{\hbar/m\omega_x}$.
 \label{Fig:Fig1}}
\end{center}
\end{figure}

The most fundamental properties of this system undergo qualitative changes across this transition.  Figs.~\ref{Fig:Fig1}(a) and (b) show examples of the condensate densities for miscible and immiscible solutions, respectively.  The component separation, shown in Fig.~\ref{Fig:Fig1}(c), is zero until $g_{12}=g_c$, above which it grows abruptly before plateauing for larger $g_{12}$.  The low-lying excitation energies, which are eigenvalues of the BdG equations, are plotted in Fig.~\ref{Fig:Fig1}(d) as a function of the interspecies interaction strength. On the miscible side ($g_{12}<g_c$), the two lowest-lying modes are out-of-phase ``slosh'' modes, which soften at the transition threshold. We perform a scaling fit $\sim |\delta|^{\nu z}$ (see Eq.~(\ref{Eq:CritScale})) to the lowest mode in the region near the transition, over $g_{12}/g_c = 0.6$-$1$, and find that $\nu z \simeq 0.505$, consistent with square root behavior.  On the immiscible side ($g_{12}>g_c$), we ignore the modes corresponding to interface bending, as they do not contribute to the transition instability~\cite{Takeuchi10}, and instead focus on the out-of-phase ``mixing'' modes~\cite{Ticknor2013,Ticknor2014}. For these modes, there is a clear deviation from a square root behavior over the fitting range $g_{12}/g_c = 1$-$1.3$, where we find $\nu z = 0.60(3)$. The uncertainty arises because of fitting ambiguity, possibly due to avoided crossings, and is particularly evident within the immediate vicinity of the transition.

\begin{figure} 
\begin{center}
\includegraphics[width=3.3in]{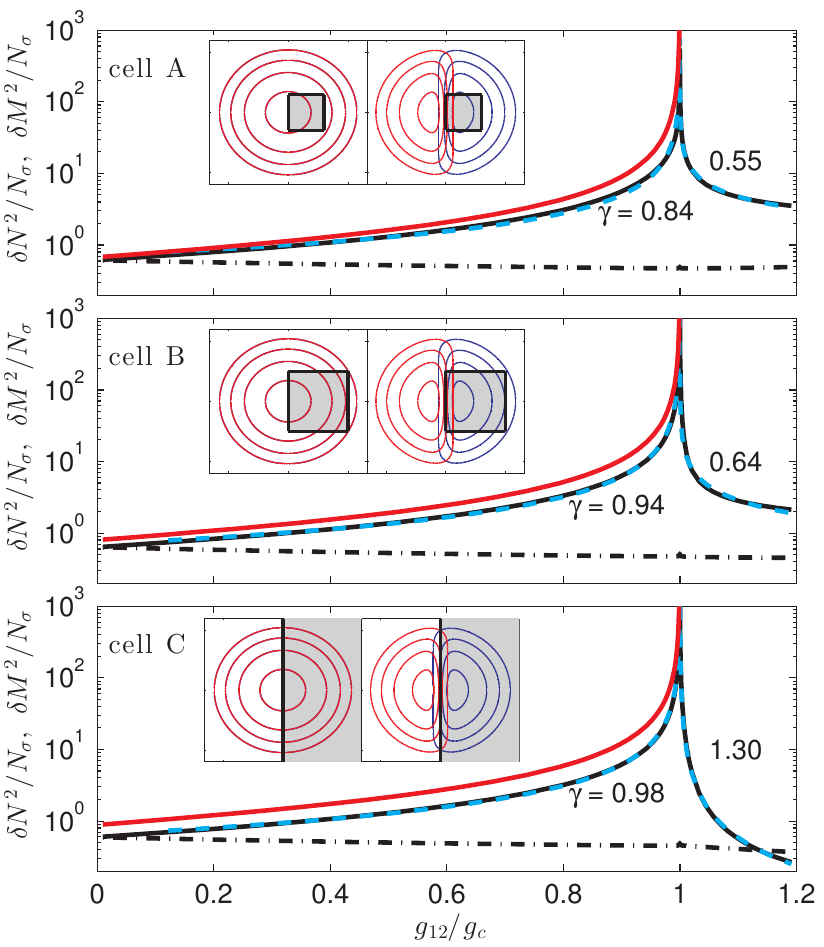}
\caption{(color online) Magnetic (solid, black) and LDA-magnetic (solid, red (gray)) number fluctuations with fits (dashed), where fitting regions are $g_{12}/g_c$ = 0.1-1 (miscible) and 1-1.2 (immiscible). Normal (dot-dashed) number fluctuations are defined as $\delta N^2 = \langle [\hat{N}_\sigma - \langle \hat{N}_\sigma \rangle ]^2 \rangle$. Cells are squares of width $L = $ $3a_x$ (cell A), $5a_x$ (cell B) and $9.3a_x$ (cell C). Insets show the cell size and position relative to the condensates for $g_{12}/g_c=0.897$ (left) and 1.094 (right). We consider a temperature $T = 7.8 \hbar \omega_x / k_B$.
 \label{Fig:FlucScaling}}
\end{center}
\end{figure}

We study number fluctuations in the low temperature regime with $T = 7.8\hbar\omega_x /k_B = T_c^0/10$, where $T_c^0 = \sqrt{6N}\hbar\omega_x /\pi k_B$ is the ideal 2D condensation temperature~\cite{Bagnato1991}.  
In Fig.~\ref{Fig:FlucScaling}, we plot number fluctuations as a function of $g_{12} / g_c$ for three square cells of various width $L$.  The cells are positioned symmetrically about the $y$-axis with their edges placed at $x=0$ (see insets).  This cell location makes them highly sensitive to fluctuations that separate (mix) the components in the miscible (immiscible) phase.  
Deep in the miscible regime ($g_{12}/g_c \ll 1$), the magnetic ($\delta M^2$) and normal ($\delta N^2$) number fluctuations are small and equal. 
The magnetic fluctuations diverge as the transition is approached from the miscible side, $g_{12}/g_c \to 1$, whereas the normal fluctuations steadily decrease. 

For the uniform binary Bose gas in the thermodynamic limit, the magnetic cell-fluctuations for the miscible phase scale as (c.f.~the susceptibility for the coherently coupled spinor system - Eq.~9 of \cite{Abad2013})
\begin{equation}\label{Eq:UniFluc}
\frac{\delta M^2}{N_1} = \frac{k_BT}{n_1 g_c\delta},
\end{equation}
where $n_1$ is the areal density of component 1, which is equal to that of component 2 in the balanced case we consider here.
Together, Eqs.~(\ref{Eq:CritScale}) and~(\ref{Eq:UniFluc}) imply that this system
has an exponent $\gamma=1$.  
Motivated by this result, we explore the scaling of magnetic fluctuations in the trapped system by fitting our numerical results to Eq.~(\ref{Eq:CritScale}).
We first perform fits over the broad range $g_{12}/g_c$ = 0.1-1, which are shown by the dashed teal lines in Fig.~\ref{Fig:FlucScaling}.   For cell A (the smallest cell) we find $\gamma =  0.84$, for cell B we find $\gamma = 0.94$, and for cell C (the largest cell, which encompasses half the system) the exponent is near unity with $\gamma=0.98$.  As the cell size is increased, the fit parameter $\gamma$ approaches unity and the  fit quality improves, i.e.~the divergence of $\delta M^2$ becomes more algebraic.
\begin{table}
\begin{center}
\begin{tabular}{| l |c|c|c|c|c|c|}
 \hline
   & \multicolumn{3}{|c|}{miscible fit regions} & \multicolumn{3}{|c|}{immiscible fit regions} \\
  \hline
 cell (width, $L$)       &  0.1-0.9 & 0.1-1   &  0.9-1 & 1-1.05  & 1-1.2    & 1.05-1.2\\ \hline
cell A ($3a_x$)    & 0.97      & 0.84     & 0.70    & 0.62     &   0.55   & 0.47\\
cell B ($5a_x$)    & 0.99      & 0.94     & 0.90    & 0.94     &  0.64    & 0.49\\
cell C ($9.3a_x$) & 1.01      & 0.98     & 0.95    & 1.29     &  1.30    & 1.24\\ \hline 
\end{tabular} 
\end{center} 
\caption{Fitting parameter $\gamma$ for magnetic fluctuations; for various ranges of $g_{12}/g_c$ and various cell sizes.} \label{Tab:Fit}
\end{table}

We also perform fits over the ranges $g_{12}/g_c=0.1$-$0.9$ and $g_{12}/g_c=0.9$-$1$, the results of which are shown in table \ref{Tab:Fit}.  An important result, is that fitting over the range $g_{12}/g_c$ = 0.1-0.9 (i.e.~excluding the transition region) produces values of $\gamma$ that are close to unity for all cell sizes.
We note that this is also the region where Bogoliubov theory is expected to be quantitatively accurate as it excludes the strongly fluctuating region.
In contrast, fitting only near the critical point ($g_{12}/g_c= 0.9$-$1$) results in $\gamma$ parameters that are further from unity, particularly for the smallest cell (cell A).
The deviation from $\gamma=1$ in smaller cells is a finite size effect, reflecting the fact that the dominant magnetic fluctuations are long-wavelength in nature; when their characteristic length (the spin healing length $\xi$) exceeds the cell size, deviations from the thermodynamic limit scaling (Eq.~(\ref{Eq:UniFluc})) are expected
\footnote{Note, similarly for the infinite uniform system, Eq.~\ref{Eq:UniFluc} is expected to hold for finite cells but only when $\xi$ is much smaller than the cell size.}.
For cell C, this cannot occur, as $\xi$ is limited by the size of the system itself.  Thus, cell C represents the closest analog to the thermodynamic limit that is achievable in the trapped system, which explains how the exponent remains close to unity when fitting to the transition region.
Additionally,  in contrast to the uniform system, the local condensate densities in the trapped system decrease with increasing $g_{12}$.  Thus, it is not surprising that the fit parameters exhibit a residual dependence on the fit range, even for cell C.  

We employ a local density approximation (LDA) by applying the uniform result (Eq.~(\ref{Eq:UniFluc})) to our numerical solutions; we replace $n_1$ with the cell-weighted average density
\begin{equation}
\bar n_{1 \sigma} = \int_\sigma d\br \frac{n^0_1(\br)}{N_{\sigma 1}^0} n_1^0(\br),
\end{equation}
where $N_{\sigma 1}^0 = \int_\sigma n_1^0(\br)d\br$, and we use the critical point for the trapped system, $g_c = 1.0072(8) g$.  It is remarkable how well the LDA results agree with the full numerical calculation, as seen in Fig.~\ref{Fig:FlucScaling}, where the LDA results are shown by the red solid lines.  Away from the transition, the LDA is more accurate when applied to cell A.  In this case, the density is more uniform across the cell (due to its small size), and the free-particle character of the magnetic fluctuations is less evident.  

As the transition is approached from the immiscible side,  the behavior is more complicated and we do not have the luxury of being able to compare our results with an unambiguous thermodynamic limit.  Again, we fit our numerical results to scaling (\ref{Eq:CritScale}) over various ranges of $g_{12}/g_c$, and show the fitting parameters $\gamma$ in table \ref{Tab:Fit}.  In this case, $\gamma$ increases significantly with increasing cell size for all fitting ranges.
Furthermore, for cells A and B, $\gamma$ is significantly larger for fits near the transition ($g_{12}/g_c = 1$-$1.05$) compared to the region further from the transition ($g_{12}/g_c = 1.05$-$1.2$).  In contrast, the fits for cell C are approximately constant over the various fitting ranges, giving $\gamma = 1.29$ near the transition, $\gamma=1.24$ away from the transition, and $\gamma = 1.30$ over the broad range $g_{12}/g_c=1$-$1.2$.
This behavior arises because the large cell (C) is insensitive to the interface bending modes, due to geometry, and is instead representative of the steeply softening mixing-modes that drive the transition [see Fig.~\ref{Fig:Fig1} (d)]. The smaller cells (A and B) however, are sensitive to both the mixing and the bending modes and consequently the scaling exponent is highly dependent on the fitting region and the relative dominance (softness) between these mode classes therein [see Fig.~\ref{Fig:Fig1} (d)].



\begin{figure} 
\begin{center}
\includegraphics[width=3.3in]{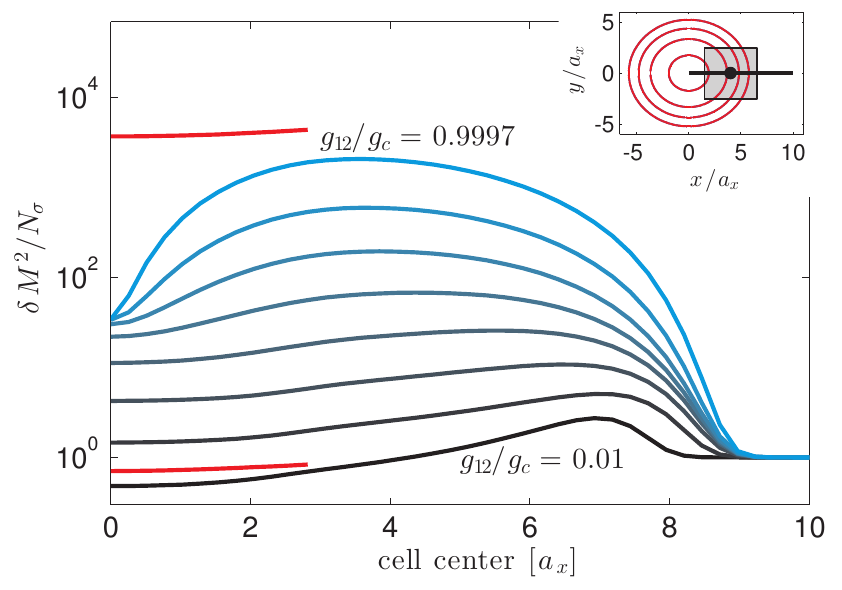}
\caption{(color online) Magnetic fluctuations vs.~cell position for a square cell with $L=5a_x$ on the miscible side. From black to cyan (gray), $g_{12}/g_c$ = 0.01, 0.63, 0.88, 0.97, 0.9905, 0.9972, 0.9991, 0.9997. The five smallest $g_{12}/g_c$ values are marked in Fig.~\ref{Fig:Fig1}(c). The LDA results for $g_{12}/g_c = 0.01$ and $0.9997$ are shown by the short red (gray) curves. Inset shows cell size and path relative to the condensate for $g_{12}/g_c = 0.897$.
 \label{Fig:FlucVsPos}}
\end{center}
\end{figure}

In Fig.~\ref{Fig:FlucVsPos}, we show how magnetic fluctuations depend on cell position in the miscible regime, with $L=5a_x$ (same size as cell B).  We calculate fluctuations for cells centered at different points along the $y=0$ axis (see inset in Fig.~\ref{Fig:FlucVsPos}).  For the case of essentially uncoupled condensates, $g_{12}/g_c = 0.01$, the behavior is similar to that of number fluctuations in a single component gas, which are subpoissonian at high density and superpoissonian at the low density condensate surface (see Fig.~5(a) of Ref.~\cite{Bisset2013}).  For cells positioned off-center, the magnetic fluctuations tend to diverge on approach to the critical point. In contrast, the fluctuations clearly converge for cells positioned at the trap center.  Here, the cells become insensitive to the out-of-phase ``slosh'' modes, since magnetic fluctuations on opposite sides correlate and nullify. Instead, the cells are most sensitive to the out-of-phase breathing modes (third and fourth-to-lowest modes in Fig.~\ref{Fig:Fig1}(d)),  which converge to a finite value at the transition.  The LDA prediction is also plotted for $g_{12}/g_c$ = 0.01 and 0.9997, shown by the solid red (gray) lines. While these results compare reasonably well with the full-numerical calculation for cell-position $\sim 3 l_x$, the LDA completely fails to capture the convergence of fluctuations at the trap center.

\section{Discussion}

The experimental realization of the system at hand could be achieved with two hyperfine states of the same atomic species \cite{Myatt1997,Hall1998} or with different species \cite{Modugno2002,McCarron2011}. As an example set of parameters, we choose $10^4$ $^{87}$Rb atoms per component and $a_{11}=a_{22}=100a_0$, where $a_0$ is the Bohr radius. Our value of $g=500\hbar^2/m$ then implies a $z$-confinement strength of $\omega_z = 2\pi \times 413$ Hz.  To observe magnetic fluctuations over two orders of magnitude on the miscible side, as we predict here, requires control of $a_{12}$ in the range 5$a_0$ to 100$a_0$.

We calculated the magnetic fluctuations at various temperatures on both sides of the transition, and found that although the magnitude depends on the temperature the scaling exponents do not, provided that the temperature is much higher than the energy of the low-lying fluctuating modes [see softening modes in Fig.~\ref{Fig:Fig1} (d)], which is easily satisfied here with $T=7.8\hbar\omega_x/k_B$.
This should be contrasted to the $T=0$ case, where fluctuations do not diverge near the transition (see e.g.~Eq.~\ref{Eq:MagFluc}).
We note that at the temperature considered, quantum fluctuations are always negligible compared to thermal fluctuations for all of our results.


\section{Summary}

We demonstrate that it is experimentally accessible to extract the exponent $\gamma$, associated with a diverging magnetic susceptibility, for the second-order miscible-immiscible transition in a trapped binary Bose gas at very low but finite temperature. Importantly, this involves \emph{in situ} density measurements, which do not require high-resolution imaging.
A  large cell that spans half the system, e.g.~the region corresponding to $x>0$, is the closest analog to the thermodynamic limit and provides the best means for extracting exponents on both sides of the transition. For the immiscible phase, such a cell is crucial to avoid contributions from the interface bending modes, which do not drive the transition.

On the miscible side of the transition, we found qualitative differences between the fully trapped and the uniform system in the thermodynamic limit, but ultimately the same scaling, $\gamma=1$, is predicted over two decades of fluctuation size for both systems. Additionally, we find the gap exponent to be $\nu z = 0.505$.
Fluctuations of the trapped system strongly depend on cell geometry and orientation, due to the dominance of  a small number of low-lying excitations.
Interestingly, fluctuations do not diverge for the special case of cells positioned at the trap center since the lowest excitations are odd functions about the origin.
On the immiscible side, we found non-trivial scaling, $\gamma  \simeq 1.30$, and an associated non-square root gap exponent, $\nu z \simeq 0.60(3)$.

We also developed and tested an LDA theory that does not require the numerically intensive calculation of Bogoliubov modes, and found that it provides reasonable agreement with the numerics in certain regimes.

\section{Acknowledgments}

We thank BV Svistunov and PB Blakie for useful discussions. RNB and CT acknowledge support from CNLS, LDRD, and LANL which is operated by LANS, LLC for the NNSA of the US DOE (contract no. DE-AC52-06NA25396).  RMW acknowledges support from an NRC postdoctoral fellowship.


\begin{thebibliography}{66}
\expandafter\ifx\csname natexlab\endcsname\relax\def\natexlab#1{#1}\fi
\expandafter\ifx\csname bibnamefont\endcsname\relax
  \def\bibnamefont#1{#1}\fi
\expandafter\ifx\csname bibfnamefont\endcsname\relax
  \def\bibfnamefont#1{#1}\fi
\expandafter\ifx\csname citenamefont\endcsname\relax
  \def\citenamefont#1{#1}\fi
\expandafter\ifx\csname url\endcsname\relax
  \def\url#1{\texttt{#1}}\fi
\expandafter\ifx\csname urlprefix\endcsname\relax\def\urlprefix{URL }\fi
\providecommand{\bibinfo}[2]{#2}
\providecommand{\eprint}[2][]{\url{#2}}

\bibitem[{\citenamefont{Anderson et~al.}(1995)\citenamefont{Anderson, Ensher,
  Matthews, Wieman, and Cornell}}]{Anderson1995}
\bibinfo{author}{\bibfnamefont{M.~H.} \bibnamefont{Anderson}},
  \bibinfo{author}{\bibfnamefont{J.~R.} \bibnamefont{Ensher}},
  \bibinfo{author}{\bibfnamefont{M.~R.} \bibnamefont{Matthews}},
  \bibinfo{author}{\bibfnamefont{C.~E.} \bibnamefont{Wieman}},
  \bibnamefont{and} \bibinfo{author}{\bibfnamefont{E.~A.}
  \bibnamefont{Cornell}}, \bibinfo{journal}{Science}
  \textbf{\bibinfo{volume}{269}}, \bibinfo{pages}{198} (\bibinfo{year}{1995}).

\bibitem[{\citenamefont{Bradley et~al.}(1995)\citenamefont{Bradley, Sackett,
  Tollett, and Hulet}}]{Bradley1995}
\bibinfo{author}{\bibfnamefont{C.~C.} \bibnamefont{Bradley}},
  \bibinfo{author}{\bibfnamefont{C.~A.} \bibnamefont{Sackett}},
  \bibinfo{author}{\bibfnamefont{J.~J.} \bibnamefont{Tollett}},
  \bibnamefont{and} \bibinfo{author}{\bibfnamefont{R.~G.} \bibnamefont{Hulet}},
  \bibinfo{journal}{Phys. Rev. Lett.} \textbf{\bibinfo{volume}{75}},
  \bibinfo{pages}{1687} (\bibinfo{year}{1995}).

\bibitem[{\citenamefont{Davis et~al.}(1995)\citenamefont{Davis, Mewes, Andrews,
  van Druten, Durfee, Kurn, and Ketterle}}]{Davis1995}
\bibinfo{author}{\bibfnamefont{K.~B.} \bibnamefont{Davis}},
  \bibinfo{author}{\bibfnamefont{M.~O.} \bibnamefont{Mewes}},
  \bibinfo{author}{\bibfnamefont{M.~R.} \bibnamefont{Andrews}},
  \bibinfo{author}{\bibfnamefont{N.~J.} \bibnamefont{van Druten}},
  \bibinfo{author}{\bibfnamefont{D.~S.} \bibnamefont{Durfee}},
  \bibinfo{author}{\bibfnamefont{D.~M.} \bibnamefont{Kurn}}, \bibnamefont{and}
  \bibinfo{author}{\bibfnamefont{W.}~\bibnamefont{Ketterle}},
  \bibinfo{journal}{Phys. Rev. Lett.} \textbf{\bibinfo{volume}{75}},
  \bibinfo{pages}{3969} (\bibinfo{year}{1995}).

\bibitem[{\citenamefont{Ensher et~al.}(1996)\citenamefont{Ensher, Jin,
  Matthews, Wieman, and Cornell}}]{Ensher1996}
\bibinfo{author}{\bibfnamefont{J.~R.} \bibnamefont{Ensher}},
  \bibinfo{author}{\bibfnamefont{D.~S.} \bibnamefont{Jin}},
  \bibinfo{author}{\bibfnamefont{M.~R.} \bibnamefont{Matthews}},
  \bibinfo{author}{\bibfnamefont{C.~E.} \bibnamefont{Wieman}},
  \bibnamefont{and} \bibinfo{author}{\bibfnamefont{E.~A.}
  \bibnamefont{Cornell}}, \bibinfo{journal}{Phys. Rev. Lett.}
  \textbf{\bibinfo{volume}{77}}, \bibinfo{pages}{4984} (\bibinfo{year}{1996}).

\bibitem[{\citenamefont{Gerbier et~al.}(2004)\citenamefont{Gerbier, Thywissen,
  Richard, Hugbart, Bouyer, and Aspect}}]{Gerbier2004}
\bibinfo{author}{\bibfnamefont{F.}~\bibnamefont{Gerbier}},
  \bibinfo{author}{\bibfnamefont{J.~H.} \bibnamefont{Thywissen}},
  \bibinfo{author}{\bibfnamefont{S.}~\bibnamefont{Richard}},
  \bibinfo{author}{\bibfnamefont{M.}~\bibnamefont{Hugbart}},
  \bibinfo{author}{\bibfnamefont{P.}~\bibnamefont{Bouyer}}, \bibnamefont{and}
  \bibinfo{author}{\bibfnamefont{A.}~\bibnamefont{Aspect}},
  \bibinfo{journal}{Phys. Rev. Lett.} \textbf{\bibinfo{volume}{92}},
  \bibinfo{pages}{030405} (\bibinfo{year}{2004}).

\bibitem[{\citenamefont{Tammuz et~al.}(2011)\citenamefont{Tammuz, Smith,
  Campbell, Beattie, Moulder, Dalibard, and Hadzibabic}}]{Tammuz2011}
\bibinfo{author}{\bibfnamefont{N.}~\bibnamefont{Tammuz}},
  \bibinfo{author}{\bibfnamefont{R.~P.} \bibnamefont{Smith}},
  \bibinfo{author}{\bibfnamefont{R.~L.~D.} \bibnamefont{Campbell}},
  \bibinfo{author}{\bibfnamefont{S.}~\bibnamefont{Beattie}},
  \bibinfo{author}{\bibfnamefont{S.}~\bibnamefont{Moulder}},
  \bibinfo{author}{\bibfnamefont{J.}~\bibnamefont{Dalibard}}, \bibnamefont{and}
  \bibinfo{author}{\bibfnamefont{Z.}~\bibnamefont{Hadzibabic}},
  \bibinfo{journal}{Phys. Rev. Lett.} \textbf{\bibinfo{volume}{106}},
  \bibinfo{pages}{230401} (\bibinfo{year}{2011}).

\bibitem[{\citenamefont{Hadzibabic et~al.}(2006)\citenamefont{Hadzibabic,
  Kr\"uger, Cheneau, Battelier, and Dalibard}}]{Hadzibabic2006}
\bibinfo{author}{\bibfnamefont{Z.}~\bibnamefont{Hadzibabic}},
  \bibinfo{author}{\bibfnamefont{P.}~\bibnamefont{Kr\"uger}},
  \bibinfo{author}{\bibfnamefont{M.}~\bibnamefont{Cheneau}},
  \bibinfo{author}{\bibfnamefont{B.}~\bibnamefont{Battelier}},
  \bibnamefont{and} \bibinfo{author}{\bibfnamefont{J.}~\bibnamefont{Dalibard}},
  \bibinfo{journal}{Nature} \textbf{\bibinfo{volume}{441}},
  \bibinfo{pages}{1118} (\bibinfo{year}{2006}).

\bibitem[{\citenamefont{Clad\'e et~al.}(2009)\citenamefont{Clad\'e, Ryu,
  Ramanathan, Helmerson, and Phillips}}]{Clade2009}
\bibinfo{author}{\bibfnamefont{P.}~\bibnamefont{Clad\'e}},
  \bibinfo{author}{\bibfnamefont{C.}~\bibnamefont{Ryu}},
  \bibinfo{author}{\bibfnamefont{A.}~\bibnamefont{Ramanathan}},
  \bibinfo{author}{\bibfnamefont{K.}~\bibnamefont{Helmerson}},
  \bibnamefont{and} \bibinfo{author}{\bibfnamefont{W.~D.}
  \bibnamefont{Phillips}}, \bibinfo{journal}{Phys. Rev. Lett.}
  \textbf{\bibinfo{volume}{102}}, \bibinfo{pages}{170401}
  (\bibinfo{year}{2009}).

\bibitem[{\citenamefont{Tung et~al.}(2010)\citenamefont{Tung, Lamporesi,
  Lobser, Xia, and Cornell}}]{Tung2010}
\bibinfo{author}{\bibfnamefont{S.}~\bibnamefont{Tung}},
  \bibinfo{author}{\bibfnamefont{G.}~\bibnamefont{Lamporesi}},
  \bibinfo{author}{\bibfnamefont{D.}~\bibnamefont{Lobser}},
  \bibinfo{author}{\bibfnamefont{L.}~\bibnamefont{Xia}}, \bibnamefont{and}
  \bibinfo{author}{\bibfnamefont{E.~A.} \bibnamefont{Cornell}},
  \bibinfo{journal}{Phys. Rev. Lett.} \textbf{\bibinfo{volume}{105}},
  \bibinfo{pages}{230408} (\bibinfo{year}{2010}).

\bibitem[{\citenamefont{Hung et~al.}(2011)\citenamefont{Hung, Zhang, Gemelke,
  and Chin}}]{Hung2011}
\bibinfo{author}{\bibfnamefont{C.-L.} \bibnamefont{Hung}},
  \bibinfo{author}{\bibfnamefont{X.}~\bibnamefont{Zhang}},
  \bibinfo{author}{\bibfnamefont{N.}~\bibnamefont{Gemelke}}, \bibnamefont{and}
  \bibinfo{author}{\bibfnamefont{C.}~\bibnamefont{Chin}},
  \bibinfo{journal}{Nature} \textbf{\bibinfo{volume}{470}},
  \bibinfo{pages}{236} (\bibinfo{year}{2011}).

\bibitem[{\citenamefont{Greiner et~al.}(2002)\citenamefont{Greiner, Mandel,
  Esslinger, H\"ansch, and Bloch}}]{Greiner2002}
\bibinfo{author}{\bibfnamefont{M.}~\bibnamefont{Greiner}},
  \bibinfo{author}{\bibfnamefont{O.}~\bibnamefont{Mandel}},
  \bibinfo{author}{\bibfnamefont{T.}~\bibnamefont{Esslinger}},
  \bibinfo{author}{\bibfnamefont{T.~W.} \bibnamefont{H\"ansch}},
  \bibnamefont{and} \bibinfo{author}{\bibfnamefont{I.}~\bibnamefont{Bloch}},
  \bibinfo{journal}{Nature} \textbf{\bibinfo{volume}{415}}, \bibinfo{pages}{39}
  (\bibinfo{year}{2002}).

\bibitem[{\citenamefont{Bakr et~al.}(2009)\citenamefont{Bakr, Gillen, Peng,
  F\"olling, and Greiner}}]{Bakr2009}
\bibinfo{author}{\bibfnamefont{W.~S.} \bibnamefont{Bakr}},
  \bibinfo{author}{\bibfnamefont{J.~I.} \bibnamefont{Gillen}},
  \bibinfo{author}{\bibfnamefont{A.}~\bibnamefont{Peng}},
  \bibinfo{author}{\bibfnamefont{S.}~\bibnamefont{F\"olling}},
  \bibnamefont{and} \bibinfo{author}{\bibfnamefont{M.}~\bibnamefont{Greiner}},
  \bibinfo{journal}{Nature} \textbf{\bibinfo{volume}{462}}, \bibinfo{pages}{74}
  (\bibinfo{year}{2009}).

\bibitem[{\citenamefont{Sherson et~al.}(2010)\citenamefont{Sherson, Weitenberg,
  Endres, Cheneau, Bloch, and Kuhr}}]{Sherson2010}
\bibinfo{author}{\bibfnamefont{J.~F.} \bibnamefont{Sherson}},
  \bibinfo{author}{\bibfnamefont{C.}~\bibnamefont{Weitenberg}},
  \bibinfo{author}{\bibfnamefont{M.}~\bibnamefont{Endres}},
  \bibinfo{author}{\bibfnamefont{M.}~\bibnamefont{Cheneau}},
  \bibinfo{author}{\bibfnamefont{I.}~\bibnamefont{Bloch}}, \bibnamefont{and}
  \bibinfo{author}{\bibfnamefont{S.}~\bibnamefont{Kuhr}},
  \bibinfo{journal}{Nature} \textbf{\bibinfo{volume}{467}}, \bibinfo{pages}{68}
  (\bibinfo{year}{2010}).

\bibitem[{\citenamefont{Bakr et~al.}(2010)\citenamefont{Bakr, Peng, Tai, Ma,
  Simon, Gillen, F\"olling, Pollet, and Greiner}}]{Bakr2010}
\bibinfo{author}{\bibfnamefont{W.~S.} \bibnamefont{Bakr}},
  \bibinfo{author}{\bibfnamefont{A.}~\bibnamefont{Peng}},
  \bibinfo{author}{\bibfnamefont{M.~E.} \bibnamefont{Tai}},
  \bibinfo{author}{\bibfnamefont{R.}~\bibnamefont{Ma}},
  \bibinfo{author}{\bibfnamefont{J.}~\bibnamefont{Simon}},
  \bibinfo{author}{\bibfnamefont{J.~I.} \bibnamefont{Gillen}},
  \bibinfo{author}{\bibfnamefont{S.}~\bibnamefont{F\"olling}},
  \bibinfo{author}{\bibfnamefont{L.}~\bibnamefont{Pollet}}, \bibnamefont{and}
  \bibinfo{author}{\bibfnamefont{M.}~\bibnamefont{Greiner}},
  \bibinfo{journal}{Science} \textbf{\bibinfo{volume}{329}},
  \bibinfo{pages}{547} (\bibinfo{year}{2010}).

\bibitem[{\citenamefont{Baumann et~al.}(2010)\citenamefont{Baumann, Guerlin,
  Brennecke, and Esslinger}}]{Baumann2010}
\bibinfo{author}{\bibfnamefont{K.}~\bibnamefont{Baumann}},
  \bibinfo{author}{\bibfnamefont{C.}~\bibnamefont{Guerlin}},
  \bibinfo{author}{\bibfnamefont{F.}~\bibnamefont{Brennecke}},
  \bibnamefont{and}
  \bibinfo{author}{\bibfnamefont{T.}~\bibnamefont{Esslinger}},
  \bibinfo{journal}{Nature} \textbf{\bibinfo{volume}{464}},
  \bibinfo{pages}{1301} (\bibinfo{year}{2010}).

\bibitem[{\citenamefont{Brennecke et~al.}(2013)\citenamefont{Brennecke, Mottl,
  Baumann, Landig, Donner, and Esslinger}}]{Brennecke2013}
\bibinfo{author}{\bibfnamefont{F.}~\bibnamefont{Brennecke}},
  \bibinfo{author}{\bibfnamefont{R.}~\bibnamefont{Mottl}},
  \bibinfo{author}{\bibfnamefont{K.}~\bibnamefont{Baumann}},
  \bibinfo{author}{\bibfnamefont{R.}~\bibnamefont{Landig}},
  \bibinfo{author}{\bibfnamefont{T.}~\bibnamefont{Donner}}, \bibnamefont{and}
  \bibinfo{author}{\bibfnamefont{T.}~\bibnamefont{Esslinger}},
  \bibinfo{journal}{Proceedings of the National Academy of Sciences}
  \textbf{\bibinfo{volume}{110}}, \bibinfo{pages}{11763}
  (\bibinfo{year}{2013}).

\bibitem[{\citenamefont{Mottl et~al.}(2012)\citenamefont{Mottl, Brennecke,
  Baumann, Landig, Donner, and Esslinger}}]{Mottl2012}
\bibinfo{author}{\bibfnamefont{R.}~\bibnamefont{Mottl}},
  \bibinfo{author}{\bibfnamefont{F.}~\bibnamefont{Brennecke}},
  \bibinfo{author}{\bibfnamefont{K.}~\bibnamefont{Baumann}},
  \bibinfo{author}{\bibfnamefont{R.}~\bibnamefont{Landig}},
  \bibinfo{author}{\bibfnamefont{T.}~\bibnamefont{Donner}}, \bibnamefont{and}
  \bibinfo{author}{\bibfnamefont{T.}~\bibnamefont{Esslinger}},
  \bibinfo{journal}{Science} \textbf{\bibinfo{volume}{336}},
  \bibinfo{pages}{1570} (\bibinfo{year}{2012}).

\bibitem[{\citenamefont{Kawaguchi and Ueda}(2012)}]{Kawaguchi2012}
\bibinfo{author}{\bibfnamefont{Y.}~\bibnamefont{Kawaguchi}} \bibnamefont{and}
  \bibinfo{author}{\bibfnamefont{M.}~\bibnamefont{Ueda}},
  \bibinfo{journal}{Physics Reports} \textbf{\bibinfo{volume}{520}},
  \bibinfo{pages}{253 } (\bibinfo{year}{2012}), ISSN \bibinfo{issn}{0370-1573},
  \bibinfo{note}{spinor Bose--Einstein condensates}.

\bibitem[{\citenamefont{Stamper-Kurn and Ueda}(2013)}]{StamperKurn2013}
\bibinfo{author}{\bibfnamefont{D.~M.} \bibnamefont{Stamper-Kurn}}
  \bibnamefont{and} \bibinfo{author}{\bibfnamefont{M.}~\bibnamefont{Ueda}},
  \bibinfo{journal}{Rev. Mod. Phys.} \textbf{\bibinfo{volume}{85}},
  \bibinfo{pages}{1191} (\bibinfo{year}{2013}).

\bibitem[{\citenamefont{Lin et~al.}(2011)\citenamefont{Lin, Jim\'enez-Garc\'ia,
  and Spielman}}]{Lin2011}
\bibinfo{author}{\bibfnamefont{Y.-J.} \bibnamefont{Lin}},
  \bibinfo{author}{\bibfnamefont{K.}~\bibnamefont{Jim\'enez-Garc\'ia}},
  \bibnamefont{and} \bibinfo{author}{\bibfnamefont{I.~B.}
  \bibnamefont{Spielman}}, \bibinfo{journal}{Nature}
  \textbf{\bibinfo{volume}{471}}, \bibinfo{pages}{83} (\bibinfo{year}{2011}).

\bibitem[{\citenamefont{Zhang et~al.}(2012)\citenamefont{Zhang, Ji, Chen,
  Zhang, Du, Yan, Pan, Zhao, Deng, Zhai et~al.}}]{Zhang2012}
\bibinfo{author}{\bibfnamefont{J.-Y.} \bibnamefont{Zhang}},
  \bibinfo{author}{\bibfnamefont{S.-C.} \bibnamefont{Ji}},
  \bibinfo{author}{\bibfnamefont{Z.}~\bibnamefont{Chen}},
  \bibinfo{author}{\bibfnamefont{L.}~\bibnamefont{Zhang}},
  \bibinfo{author}{\bibfnamefont{Z.-D.} \bibnamefont{Du}},
  \bibinfo{author}{\bibfnamefont{B.}~\bibnamefont{Yan}},
  \bibinfo{author}{\bibfnamefont{G.-S.} \bibnamefont{Pan}},
  \bibinfo{author}{\bibfnamefont{B.}~\bibnamefont{Zhao}},
  \bibinfo{author}{\bibfnamefont{Y.-J.} \bibnamefont{Deng}},
  \bibinfo{author}{\bibfnamefont{H.}~\bibnamefont{Zhai}}, \bibnamefont{et~al.},
  \bibinfo{journal}{Phys. Rev. Lett.} \textbf{\bibinfo{volume}{109}},
  \bibinfo{pages}{115301} (\bibinfo{year}{2012}).

\bibitem[{\citenamefont{Parker et~al.}(2013)\citenamefont{Parker, Ha, and
  Chin}}]{Parker2013}
\bibinfo{author}{\bibfnamefont{C.~V.} \bibnamefont{Parker}},
  \bibinfo{author}{\bibfnamefont{L.-C.} \bibnamefont{Ha}}, \bibnamefont{and}
  \bibinfo{author}{\bibfnamefont{C.}~\bibnamefont{Chin}},
  \bibinfo{journal}{Nature Physics} \textbf{\bibinfo{volume}{9}},
  \bibinfo{pages}{769} (\bibinfo{year}{2013}).

\bibitem[{\citenamefont{Papp et~al.}(2008)\citenamefont{Papp, Pino, and
  Wieman}}]{Papp2008}
\bibinfo{author}{\bibfnamefont{S.~B.} \bibnamefont{Papp}},
  \bibinfo{author}{\bibfnamefont{J.~M.} \bibnamefont{Pino}}, \bibnamefont{and}
  \bibinfo{author}{\bibfnamefont{C.~E.} \bibnamefont{Wieman}},
  \bibinfo{journal}{Phys. Rev. Lett.} \textbf{\bibinfo{volume}{101}},
  \bibinfo{pages}{040402} (\bibinfo{year}{2008}).

\bibitem[{\citenamefont{Hall et~al.}(1998{\natexlab{a}})\citenamefont{Hall,
  Matthews, Ensher, Wieman, and Cornell}}]{Hall1998B}
\bibinfo{author}{\bibfnamefont{D.~S.} \bibnamefont{Hall}},
  \bibinfo{author}{\bibfnamefont{M.~R.} \bibnamefont{Matthews}},
  \bibinfo{author}{\bibfnamefont{J.~R.} \bibnamefont{Ensher}},
  \bibinfo{author}{\bibfnamefont{C.~E.} \bibnamefont{Wieman}},
  \bibnamefont{and} \bibinfo{author}{\bibfnamefont{E.~A.}
  \bibnamefont{Cornell}}, \bibinfo{journal}{Phys. Rev. Lett.}
  \textbf{\bibinfo{volume}{81}}, \bibinfo{pages}{1539}
  (\bibinfo{year}{1998}{\natexlab{a}}).

\bibitem[{\citenamefont{Esry et~al.}(1997)\citenamefont{Esry, Greene, Burke,
  and Bohn}}]{Esry1997}
\bibinfo{author}{\bibfnamefont{B.~D.} \bibnamefont{Esry}},
  \bibinfo{author}{\bibfnamefont{C.~H.} \bibnamefont{Greene}},
  \bibinfo{author}{\bibfnamefont{J.~P.} \bibnamefont{Burke},
  \bibfnamefont{Jr.}}, \bibnamefont{and} \bibinfo{author}{\bibfnamefont{J.~L.}
  \bibnamefont{Bohn}}, \bibinfo{journal}{Phys. Rev. Lett.}
  \textbf{\bibinfo{volume}{78}}, \bibinfo{pages}{3594} (\bibinfo{year}{1997}).

\bibitem[{\citenamefont{Riboli and Modugno}(2002)}]{Riboli2002}
\bibinfo{author}{\bibfnamefont{F.}~\bibnamefont{Riboli}} \bibnamefont{and}
  \bibinfo{author}{\bibfnamefont{M.}~\bibnamefont{Modugno}},
  \bibinfo{journal}{Phys. Rev. A} \textbf{\bibinfo{volume}{65}},
  \bibinfo{pages}{063614} (\bibinfo{year}{2002}).

\bibitem[{\citenamefont{Svidzinsky and Chui}(2003)}]{Svidzinsky2003}
\bibinfo{author}{\bibfnamefont{A.~A.} \bibnamefont{Svidzinsky}}
  \bibnamefont{and} \bibinfo{author}{\bibfnamefont{S.~T.} \bibnamefont{Chui}},
  \bibinfo{journal}{Phys. Rev. A} \textbf{\bibinfo{volume}{67}},
  \bibinfo{pages}{053608} (\bibinfo{year}{2003}).

\bibitem[{\citenamefont{Trippenbach et~al.}(2000)\citenamefont{Trippenbach,
  G\'oral, Rzazewski, Malomed, and Band}}]{Trippenbach2000}
\bibinfo{author}{\bibfnamefont{M.}~\bibnamefont{Trippenbach}},
  \bibinfo{author}{\bibfnamefont{K.}~\bibnamefont{G\'oral}},
  \bibinfo{author}{\bibfnamefont{K.}~\bibnamefont{Rzazewski}},
  \bibinfo{author}{\bibfnamefont{B.}~\bibnamefont{Malomed}}, \bibnamefont{and}
  \bibinfo{author}{\bibfnamefont{Y.~B.} \bibnamefont{Band}},
  \bibinfo{journal}{Journal of Physics B: Atomic, Molecular and Optical
  Physics} \textbf{\bibinfo{volume}{33}}, \bibinfo{pages}{4017}
  (\bibinfo{year}{2000}).

\bibitem[{\citenamefont{Timmermans}(1998)}]{Timmermans1998}
\bibinfo{author}{\bibfnamefont{E.}~\bibnamefont{Timmermans}},
  \bibinfo{journal}{Phys. Rev. Lett.} \textbf{\bibinfo{volume}{81}},
  \bibinfo{pages}{5718} (\bibinfo{year}{1998}).

\bibitem[{\citenamefont{Ao and Chui}(1998)}]{Ao1998}
\bibinfo{author}{\bibfnamefont{P.}~\bibnamefont{Ao}} \bibnamefont{and}
  \bibinfo{author}{\bibfnamefont{S.~T.} \bibnamefont{Chui}},
  \bibinfo{journal}{Phys. Rev. A} \textbf{\bibinfo{volume}{58}},
  \bibinfo{pages}{4836} (\bibinfo{year}{1998}).

\bibitem[{\citenamefont{Ao and Chui}(2000)}]{Ao2000}
\bibinfo{author}{\bibfnamefont{P.}~\bibnamefont{Ao}} \bibnamefont{and}
  \bibinfo{author}{\bibfnamefont{S.~T.} \bibnamefont{Chui}},
  \bibinfo{journal}{Journal of Physics B: Atomic, Molecular and Optical
  Physics} \textbf{\bibinfo{volume}{33}}, \bibinfo{pages}{535}
  (\bibinfo{year}{2000}).

\bibitem[{\citenamefont{Alexandrov and Kabanov}(2002)}]{Alexandrov2002}
\bibinfo{author}{\bibfnamefont{A.~S.} \bibnamefont{Alexandrov}}
  \bibnamefont{and} \bibinfo{author}{\bibfnamefont{V.~V.}
  \bibnamefont{Kabanov}}, \bibinfo{journal}{Journal of Physics: Condensed
  Matter} \textbf{\bibinfo{volume}{14}}, \bibinfo{pages}{L327}
  (\bibinfo{year}{2002}).

\bibitem[{\citenamefont{Santamore and Timmermans}(2012)}]{Santamore2012}
\bibinfo{author}{\bibfnamefont{D.~H.} \bibnamefont{Santamore}}
  \bibnamefont{and}
  \bibinfo{author}{\bibfnamefont{E.}~\bibnamefont{Timmermans}},
  \bibinfo{journal}{Europhysics Letters} \textbf{\bibinfo{volume}{97}},
  \bibinfo{pages}{36009} (\bibinfo{year}{2012}).

\bibitem[{\citenamefont{Ole\'s and Sacha}(2008)}]{Oles2008}
\bibinfo{author}{\bibfnamefont{B.}~\bibnamefont{Ole\'s}} \bibnamefont{and}
  \bibinfo{author}{\bibfnamefont{K.}~\bibnamefont{Sacha}},
  \bibinfo{journal}{Journal of Physics A: Mathematical and Theoretical}
  \textbf{\bibinfo{volume}{41}}, \bibinfo{pages}{145005}
  (\bibinfo{year}{2008}).

\bibitem[{\citenamefont{Zhan et~al.}(2014)\citenamefont{Zhan, Sabbatini, Davis,
  and McCulloch}}]{Zhan2014}
\bibinfo{author}{\bibfnamefont{F.}~\bibnamefont{Zhan}},
  \bibinfo{author}{\bibfnamefont{J.}~\bibnamefont{Sabbatini}},
  \bibinfo{author}{\bibfnamefont{M.~J.} \bibnamefont{Davis}}, \bibnamefont{and}
  \bibinfo{author}{\bibfnamefont{I.~P.} \bibnamefont{McCulloch}},
  \bibinfo{journal}{Phys. Rev. A} \textbf{\bibinfo{volume}{90}},
  \bibinfo{pages}{023630} (\bibinfo{year}{2014}).

\bibitem[{\citenamefont{Pattinson et~al.}(2013)\citenamefont{Pattinson, Billam,
  Gardiner, McCarron, Cho, Cornish, Parker, and Proukakis}}]{Pattinson2013}
\bibinfo{author}{\bibfnamefont{R.~W.} \bibnamefont{Pattinson}},
  \bibinfo{author}{\bibfnamefont{T.~P.} \bibnamefont{Billam}},
  \bibinfo{author}{\bibfnamefont{S.~A.} \bibnamefont{Gardiner}},
  \bibinfo{author}{\bibfnamefont{D.~J.} \bibnamefont{McCarron}},
  \bibinfo{author}{\bibfnamefont{H.~W.} \bibnamefont{Cho}},
  \bibinfo{author}{\bibfnamefont{S.~L.} \bibnamefont{Cornish}},
  \bibinfo{author}{\bibfnamefont{N.~G.} \bibnamefont{Parker}},
  \bibnamefont{and} \bibinfo{author}{\bibfnamefont{N.~P.}
  \bibnamefont{Proukakis}}, \bibinfo{journal}{Phys. Rev. A}
  \textbf{\bibinfo{volume}{87}}, \bibinfo{pages}{013625}
  (\bibinfo{year}{2013}).

\bibitem[{\citenamefont{Roy and Angom}(2014)}]{Roy2014}
\bibinfo{author}{\bibfnamefont{A.}~\bibnamefont{Roy}} \bibnamefont{and}
  \bibinfo{author}{\bibfnamefont{D.}~\bibnamefont{Angom}},
  \bibinfo{journal}{Phys. Rev. A} \textbf{\bibinfo{volume}{90}},
  \bibinfo{pages}{023612} (\bibinfo{year}{2014}).

\bibitem[{\citenamefont{Mertes et~al.}(2007)\citenamefont{Mertes, Merrill,
  Carretero-Gonz\'alez, Frantzeskakis, Kevrekidis, and Hall}}]{Mertes2007}
\bibinfo{author}{\bibfnamefont{K.~M.} \bibnamefont{Mertes}},
  \bibinfo{author}{\bibfnamefont{J.~W.} \bibnamefont{Merrill}},
  \bibinfo{author}{\bibfnamefont{R.}~\bibnamefont{Carretero-Gonz\'alez}},
  \bibinfo{author}{\bibfnamefont{D.~J.} \bibnamefont{Frantzeskakis}},
  \bibinfo{author}{\bibfnamefont{P.~G.} \bibnamefont{Kevrekidis}},
  \bibnamefont{and} \bibinfo{author}{\bibfnamefont{D.~S.} \bibnamefont{Hall}},
  \bibinfo{journal}{Phys. Rev. Lett.} \textbf{\bibinfo{volume}{99}},
  \bibinfo{pages}{190402} (\bibinfo{year}{2007}).

\bibitem[{\citenamefont{Anderson et~al.}(2009)\citenamefont{Anderson, Ticknor,
  Sidorov, and Hall}}]{Anderson2009}
\bibinfo{author}{\bibfnamefont{R.~P.} \bibnamefont{Anderson}},
  \bibinfo{author}{\bibfnamefont{C.}~\bibnamefont{Ticknor}},
  \bibinfo{author}{\bibfnamefont{A.~I.} \bibnamefont{Sidorov}},
  \bibnamefont{and} \bibinfo{author}{\bibfnamefont{B.~V.} \bibnamefont{Hall}},
  \bibinfo{journal}{Phys. Rev. A} \textbf{\bibinfo{volume}{80}},
  \bibinfo{pages}{023603} (\bibinfo{year}{2009}).

\bibitem[{\citenamefont{Kasamatsu and Tsubota}(2004)}]{Kasamatsu2004}
\bibinfo{author}{\bibfnamefont{K.}~\bibnamefont{Kasamatsu}} \bibnamefont{and}
  \bibinfo{author}{\bibfnamefont{M.}~\bibnamefont{Tsubota}},
  \bibinfo{journal}{Phys. Rev. Lett.} \textbf{\bibinfo{volume}{93}},
  \bibinfo{pages}{100402} (\bibinfo{year}{2004}).

\bibitem[{\citenamefont{Ronen et~al.}(2008)\citenamefont{Ronen, Bohn, Halmo,
  and Edwards}}]{Ronen2008}
\bibinfo{author}{\bibfnamefont{S.}~\bibnamefont{Ronen}},
  \bibinfo{author}{\bibfnamefont{J.~L.} \bibnamefont{Bohn}},
  \bibinfo{author}{\bibfnamefont{L.~E.} \bibnamefont{Halmo}}, \bibnamefont{and}
  \bibinfo{author}{\bibfnamefont{M.}~\bibnamefont{Edwards}},
  \bibinfo{journal}{Phys. Rev. A} \textbf{\bibinfo{volume}{78}},
  \bibinfo{pages}{053613} (\bibinfo{year}{2008}).

\bibitem[{\citenamefont{Hofmann et~al.}(2014)\citenamefont{Hofmann, Natu, and
  Das~Sarma}}]{Hofmann2014}
\bibinfo{author}{\bibfnamefont{J.}~\bibnamefont{Hofmann}},
  \bibinfo{author}{\bibfnamefont{S.~S.} \bibnamefont{Natu}}, \bibnamefont{and}
  \bibinfo{author}{\bibfnamefont{S.}~\bibnamefont{Das~Sarma}},
  \bibinfo{journal}{Phys. Rev. Lett.} \textbf{\bibinfo{volume}{113}},
  \bibinfo{pages}{095702} (\bibinfo{year}{2014}).

\bibitem[{\citenamefont{Nicklas et~al.}(2014)\citenamefont{Nicklas, Muessel,
  Strobel, Kevrekidis, and Oberthaler}}]{Nicklas2014}
\bibinfo{author}{\bibfnamefont{E.}~\bibnamefont{Nicklas}},
  \bibinfo{author}{\bibfnamefont{W.}~\bibnamefont{Muessel}},
  \bibinfo{author}{\bibfnamefont{H.}~\bibnamefont{Strobel}},
  \bibinfo{author}{\bibfnamefont{P.}~\bibnamefont{Kevrekidis}},
  \bibnamefont{and}
  \bibinfo{author}{\bibfnamefont{M.}~\bibnamefont{Oberthaler}},
  \bibinfo{journal}{ArXiv e-prints}  (\bibinfo{year}{2014}),
  \eprint{1407.8049}.

\bibitem[{\citenamefont{Sabbatini et~al.}(2011)\citenamefont{Sabbatini, Zurek,
  and Davis}}]{Sabbatini2011}
\bibinfo{author}{\bibfnamefont{J.}~\bibnamefont{Sabbatini}},
  \bibinfo{author}{\bibfnamefont{W.~H.} \bibnamefont{Zurek}}, \bibnamefont{and}
  \bibinfo{author}{\bibfnamefont{M.~J.} \bibnamefont{Davis}},
  \bibinfo{journal}{Phys. Rev. Lett.} \textbf{\bibinfo{volume}{107}},
  \bibinfo{pages}{230402} (\bibinfo{year}{2011}).

\bibitem[{\citenamefont{Jacqmin et~al.}(2011)\citenamefont{Jacqmin, Armijo,
  Berrada, Kheruntsyan, and Bouchoule}}]{Jacqmin2011}
\bibinfo{author}{\bibfnamefont{T.}~\bibnamefont{Jacqmin}},
  \bibinfo{author}{\bibfnamefont{J.}~\bibnamefont{Armijo}},
  \bibinfo{author}{\bibfnamefont{T.}~\bibnamefont{Berrada}},
  \bibinfo{author}{\bibfnamefont{K.~V.} \bibnamefont{Kheruntsyan}},
  \bibnamefont{and}
  \bibinfo{author}{\bibfnamefont{I.}~\bibnamefont{Bouchoule}},
  \bibinfo{journal}{Phys. Rev. Lett.} \textbf{\bibinfo{volume}{106}},
  \bibinfo{pages}{230405} (\bibinfo{year}{2011}).

\bibitem[{\citenamefont{Blumkin et~al.}(2013)\citenamefont{Blumkin, Rinott,
  Schley, Berkovitz, Shammass, and Steinhauer}}]{Blumkin2013}
\bibinfo{author}{\bibfnamefont{A.}~\bibnamefont{Blumkin}},
  \bibinfo{author}{\bibfnamefont{S.}~\bibnamefont{Rinott}},
  \bibinfo{author}{\bibfnamefont{R.}~\bibnamefont{Schley}},
  \bibinfo{author}{\bibfnamefont{A.}~\bibnamefont{Berkovitz}},
  \bibinfo{author}{\bibfnamefont{I.}~\bibnamefont{Shammass}}, \bibnamefont{and}
  \bibinfo{author}{\bibfnamefont{J.}~\bibnamefont{Steinhauer}},
  \bibinfo{journal}{Phys. Rev. Lett.} \textbf{\bibinfo{volume}{110}},
  \bibinfo{pages}{265301} (\bibinfo{year}{2013}).

\bibitem[{\citenamefont{Symes et~al.}(2014)\citenamefont{Symes, Baillie, and
  Blakie}}]{Symes2014}
\bibinfo{author}{\bibfnamefont{L.~M.} \bibnamefont{Symes}},
  \bibinfo{author}{\bibfnamefont{D.}~\bibnamefont{Baillie}}, \bibnamefont{and}
  \bibinfo{author}{\bibfnamefont{P.~B.} \bibnamefont{Blakie}},
  \bibinfo{journal}{Phys. Rev. A} \textbf{\bibinfo{volume}{90}},
  \bibinfo{pages}{053616} (\bibinfo{year}{2014}).

\bibitem[{\citenamefont{Bisset and Blakie}(2013)}]{Bisset2013B}
\bibinfo{author}{\bibfnamefont{R.~N.} \bibnamefont{Bisset}} \bibnamefont{and}
  \bibinfo{author}{\bibfnamefont{P.~B.} \bibnamefont{Blakie}},
  \bibinfo{journal}{Phys. Rev. Lett.} \textbf{\bibinfo{volume}{110}},
  \bibinfo{pages}{265302} (\bibinfo{year}{2013}).

\bibitem[{\citenamefont{Bisset et~al.}(2013)\citenamefont{Bisset, Ticknor, and
  Blakie}}]{Bisset2013}
\bibinfo{author}{\bibfnamefont{R.~N.} \bibnamefont{Bisset}},
  \bibinfo{author}{\bibfnamefont{C.}~\bibnamefont{Ticknor}}, \bibnamefont{and}
  \bibinfo{author}{\bibfnamefont{P.~B.} \bibnamefont{Blakie}},
  \bibinfo{journal}{Phys. Rev. A} \textbf{\bibinfo{volume}{88}},
  \bibinfo{pages}{063624} (\bibinfo{year}{2013}).

\bibitem[{\citenamefont{Klawunn et~al.}(2011)\citenamefont{Klawunn, Recati,
  Pitaevskii, and Stringari}}]{Klawunn2011}
\bibinfo{author}{\bibfnamefont{M.}~\bibnamefont{Klawunn}},
  \bibinfo{author}{\bibfnamefont{A.}~\bibnamefont{Recati}},
  \bibinfo{author}{\bibfnamefont{L.~P.} \bibnamefont{Pitaevskii}},
  \bibnamefont{and}
  \bibinfo{author}{\bibfnamefont{S.}~\bibnamefont{Stringari}},
  \bibinfo{journal}{Phys. Rev. A} \textbf{\bibinfo{volume}{84}},
  \bibinfo{pages}{033612} (\bibinfo{year}{2011}).

\bibitem[{\citenamefont{Sachdev}(2011)}]{SachdevBook}
\bibinfo{author}{\bibfnamefont{S.}~\bibnamefont{Sachdev}},
  \emph{\bibinfo{title}{{Quantum Phase Transitions}}}
  (\bibinfo{publisher}{Cambridge University Press}, \bibinfo{year}{2011}),
  \bibinfo{edition}{2nd} ed.

\bibitem[{\citenamefont{Petrov et~al.}(2000)\citenamefont{Petrov, Holzmann, and
  Shlyapnikov}}]{Petrov00}
\bibinfo{author}{\bibfnamefont{D.~S.} \bibnamefont{Petrov}},
  \bibinfo{author}{\bibfnamefont{M.}~\bibnamefont{Holzmann}}, \bibnamefont{and}
  \bibinfo{author}{\bibfnamefont{G.~V.} \bibnamefont{Shlyapnikov}},
  \bibinfo{journal}{Phys. Rev. Lett.} \textbf{\bibinfo{volume}{84}},
  \bibinfo{pages}{2551} (\bibinfo{year}{2000}).

\bibitem[{\citenamefont{Pu and Bigelow}(1998{\natexlab{a}})}]{Pu1998}
\bibinfo{author}{\bibfnamefont{H.}~\bibnamefont{Pu}} \bibnamefont{and}
  \bibinfo{author}{\bibfnamefont{N.~P.} \bibnamefont{Bigelow}},
  \bibinfo{journal}{Phys. Rev. Lett.} \textbf{\bibinfo{volume}{80}},
  \bibinfo{pages}{1130} (\bibinfo{year}{1998}{\natexlab{a}}).

\bibitem[{\citenamefont{Ho and Shenoy}(1996)}]{Ho1996}
\bibinfo{author}{\bibfnamefont{T.-L.} \bibnamefont{Ho}} \bibnamefont{and}
  \bibinfo{author}{\bibfnamefont{V.~B.} \bibnamefont{Shenoy}},
  \bibinfo{journal}{Phys. Rev. Lett.} \textbf{\bibinfo{volume}{77}},
  \bibinfo{pages}{3276} (\bibinfo{year}{1996}).

\bibitem[{\citenamefont{Ticknor}(2013)}]{Ticknor2013}
\bibinfo{author}{\bibfnamefont{C.}~\bibnamefont{Ticknor}},
  \bibinfo{journal}{Phys. Rev. A} \textbf{\bibinfo{volume}{88}},
  \bibinfo{pages}{013623} (\bibinfo{year}{2013}).

\bibitem[{\citenamefont{Pu and Bigelow}(1998{\natexlab{b}})}]{Pu1998A}
\bibinfo{author}{\bibfnamefont{H.}~\bibnamefont{Pu}} \bibnamefont{and}
  \bibinfo{author}{\bibfnamefont{N.~P.} \bibnamefont{Bigelow}},
  \bibinfo{journal}{Phys. Rev. Lett.} \textbf{\bibinfo{volume}{80}},
  \bibinfo{pages}{1134} (\bibinfo{year}{1998}{\natexlab{b}}).

\bibitem[{\citenamefont{Recati and Stringari}(2011)}]{Recati2011}
\bibinfo{author}{\bibfnamefont{A.}~\bibnamefont{Recati}} \bibnamefont{and}
  \bibinfo{author}{\bibfnamefont{S.}~\bibnamefont{Stringari}},
  \bibinfo{journal}{Phys. Rev. Lett.} \textbf{\bibinfo{volume}{106}},
  \bibinfo{pages}{080402} (\bibinfo{year}{2011}).

\bibitem[{\citenamefont{Seo and de~Melo}(2011)}]{Seo2011}
\bibinfo{author}{\bibfnamefont{K.}~\bibnamefont{Seo}} \bibnamefont{and}
  \bibinfo{author}{\bibfnamefont{C.~A. R.~S.} \bibnamefont{de~Melo}},
  \bibinfo{journal}{ArXiv e-prints}  (\bibinfo{year}{2011}),
  \eprint{1105.4365}.

\bibitem[{\citenamefont{Takeuchi et~al.}(2010)\citenamefont{Takeuchi, Suzuki,
  Kasamatsu, Saito, and Tsubota}}]{Takeuchi10}
\bibinfo{author}{\bibfnamefont{H.}~\bibnamefont{Takeuchi}},
  \bibinfo{author}{\bibfnamefont{N.}~\bibnamefont{Suzuki}},
  \bibinfo{author}{\bibfnamefont{K.}~\bibnamefont{Kasamatsu}},
  \bibinfo{author}{\bibfnamefont{H.}~\bibnamefont{Saito}}, \bibnamefont{and}
  \bibinfo{author}{\bibfnamefont{M.}~\bibnamefont{Tsubota}},
  \bibinfo{journal}{Phys. Rev. B} \textbf{\bibinfo{volume}{81}},
  \bibinfo{pages}{094517} (\bibinfo{year}{2010}).

\bibitem[{\citenamefont{Ticknor}(2014)}]{Ticknor2014}
\bibinfo{author}{\bibfnamefont{C.}~\bibnamefont{Ticknor}},
  \bibinfo{journal}{Phys. Rev. A} \textbf{\bibinfo{volume}{89}},
  \bibinfo{pages}{053601} (\bibinfo{year}{2014}).

\bibitem[{\citenamefont{Bagnato and Kleppner}(1991)}]{Bagnato1991}
\bibinfo{author}{\bibfnamefont{V.}~\bibnamefont{Bagnato}} \bibnamefont{and}
  \bibinfo{author}{\bibfnamefont{D.}~\bibnamefont{Kleppner}},
  \bibinfo{journal}{Phys. Rev. A} \textbf{\bibinfo{volume}{44}},
  \bibinfo{pages}{7439} (\bibinfo{year}{1991}).

\bibitem[{\citenamefont{Abad and Recati}(2013)}]{Abad2013}
\bibinfo{author}{\bibfnamefont{M.}~\bibnamefont{Abad}} \bibnamefont{and}
  \bibinfo{author}{\bibfnamefont{A.}~\bibnamefont{Recati}},
  \bibinfo{journal}{The European Physical Journal D}
  \textbf{\bibinfo{volume}{67}}, \bibinfo{eid}{148} (\bibinfo{year}{2013}),
  ISSN \bibinfo{issn}{1434-6060}.

\bibitem[{\citenamefont{Myatt et~al.}(1997)\citenamefont{Myatt, Burt, Ghrist,
  Cornell, and Wieman}}]{Myatt1997}
\bibinfo{author}{\bibfnamefont{C.~J.} \bibnamefont{Myatt}},
  \bibinfo{author}{\bibfnamefont{E.~A.} \bibnamefont{Burt}},
  \bibinfo{author}{\bibfnamefont{R.~W.} \bibnamefont{Ghrist}},
  \bibinfo{author}{\bibfnamefont{E.~A.} \bibnamefont{Cornell}},
  \bibnamefont{and} \bibinfo{author}{\bibfnamefont{C.~E.}
  \bibnamefont{Wieman}}, \bibinfo{journal}{Phys. Rev. Lett.}
  \textbf{\bibinfo{volume}{78}}, \bibinfo{pages}{586} (\bibinfo{year}{1997}).

\bibitem[{\citenamefont{Hall et~al.}(1998{\natexlab{b}})\citenamefont{Hall,
  Matthews, Wieman, and Cornell}}]{Hall1998}
\bibinfo{author}{\bibfnamefont{D.~S.} \bibnamefont{Hall}},
  \bibinfo{author}{\bibfnamefont{M.~R.} \bibnamefont{Matthews}},
  \bibinfo{author}{\bibfnamefont{C.~E.} \bibnamefont{Wieman}},
  \bibnamefont{and} \bibinfo{author}{\bibfnamefont{E.~A.}
  \bibnamefont{Cornell}}, \bibinfo{journal}{Phys. Rev. Lett.}
  \textbf{\bibinfo{volume}{81}}, \bibinfo{pages}{1543}
  (\bibinfo{year}{1998}{\natexlab{b}}).

\bibitem[{\citenamefont{Modugno et~al.}(2002)\citenamefont{Modugno, Modugno,
  Riboli, Roati, and Inguscio}}]{Modugno2002}
\bibinfo{author}{\bibfnamefont{G.}~\bibnamefont{Modugno}},
  \bibinfo{author}{\bibfnamefont{M.}~\bibnamefont{Modugno}},
  \bibinfo{author}{\bibfnamefont{F.}~\bibnamefont{Riboli}},
  \bibinfo{author}{\bibfnamefont{G.}~\bibnamefont{Roati}}, \bibnamefont{and}
  \bibinfo{author}{\bibfnamefont{M.}~\bibnamefont{Inguscio}},
  \bibinfo{journal}{Phys. Rev. Lett.} \textbf{\bibinfo{volume}{89}},
  \bibinfo{pages}{190404} (\bibinfo{year}{2002}).

\bibitem[{\citenamefont{McCarron et~al.}(2011)\citenamefont{McCarron, Cho,
  Jenkin, K\"oppinger, and Cornish}}]{McCarron2011}
\bibinfo{author}{\bibfnamefont{D.~J.} \bibnamefont{McCarron}},
  \bibinfo{author}{\bibfnamefont{H.~W.} \bibnamefont{Cho}},
  \bibinfo{author}{\bibfnamefont{D.~L.} \bibnamefont{Jenkin}},
  \bibinfo{author}{\bibfnamefont{M.~P.} \bibnamefont{K\"oppinger}},
  \bibnamefont{and} \bibinfo{author}{\bibfnamefont{S.~L.}
  \bibnamefont{Cornish}}, \bibinfo{journal}{Phys. Rev. A}
  \textbf{\bibinfo{volume}{84}}, \bibinfo{pages}{011603}
  (\bibinfo{year}{2011}).

\end{thebibliography}

\end{document}